# Bloch wave filtering in tetrachiral materials via mechanical tuning

F Vadalà[1], A Bacigalupo[2], M Lepidi[1] and L Gambarotta[1]

[1]DICCA, University of Genoa, Via Montallegro 1, 16145, Genova, Italy,

[2]IMT School for Advanced Studies, Piazza S. Francesco 19, 55100, Lucca, Italy.

E-mail:  francesca.vadala@edu.unige.it, andrea.bacigalupo@imtlucca.it,
marco.lepidi@unige.it, luigi.gambarotta@unige.it.

**Abstract.** The periodic cellular topology characterizing the microscale structure of a heterogeneous material may allow the finest functional customization of its acoustic dispersion properties. The paper addresses the free propagation of elastic waves in micro-structured cellular materials. Focus is on the alternative formulations suited to describe the wave propagation in the material, according to the classic canons of solid or structural mechanics. Adopting the centrosymmetric tetrachiral microstructure as prototypical periodic cell, the frequency dispersion spectrum resulting from a synthetic lagrangian beam-lattice formulation is compared with its counterpart derived from different continuous models (high-fidelity first-order heterogeneous and equivalent homogenized micropolar continuum). Asymptotic perturbation-based approximations and numerical spectral solutions are cross-validated. Adopting the low-frequency band gaps of the material band structures as functional targets, parametric analyses are carried out to highlight the descriptive limits of the synthetic models and to explore the enlarged parameter space described by high-fidelity models. The final tuning of the mechanical properties of the cellular microstructure is employed to successfully verify the wave filtering functionality of the tetrachiral material.

Corresponding authors: Francesca Vadalà, mail: francesca.vadala@edu.unige.it
Andrea Bacigalupo, mail: andrea.bacigalupo@imtlucca.it
Marco Lepidi, mail: marco.lepidi@unige.it

# 1. Introduction

Periodic materials are characterized by a repetitive microstructure realizing a regular pattern of elementary cells. The research interest in these materials is being currently renewed for their high mechanical performances and smart technological applications in the naval, aerospace, nuclear, sport, biomedical engineering fields. The key of such an exponential success can be recognized in their non-conventional, or even extreme mechanical properties and tunable multi-purpose functionalities [1],[2].

Within the wide realm of microstructured periodic materials, two leading research lines can be identified. The first line pays attention to the homogenization or continualization in local and nonlocal continua in which the overall constitutive tensors are determined by means of standard or generalized macro-homogeneity conditions [3],[4],[5],[6],[7],[8],[9],[10]. The second line focuses on the assessment and customization of the acoustic dispersion properties associated to the propagation of Bloch waves across the material, either in its original periodic microstructure [11],[12],[13],[14] or in its equivalent homogenized form [8],[9],[15],[16],[17]. In this respect, the periodic materials with a chiral or antichiral microstructure of the elementary cell [18],[19],[20], consisting of stiff disks or rings, connected by a variable number of flexible ligaments, are particularly attractive for their potential as acoustic waveguides or phononic filters. In the current literature dealing with this material class, the pass and stop bands characterizing the band structures have been determined by solving the dispersion problem related to low-dimensional lagrangian models [21],[22],[23],[24],[25],[26] high-fidelity micromechanical formulations accounting for the material heterogeneity at the microscale [11],[12],[14],[27] and equivalent local and non-local homogenized continua [9],[24],[25],[26]. The underlying idea is that, within certain physically admissible ranges, the geometric and mechanical parameters can be intended as freely tunable variables for customizing the acoustic dispersion properties of the material. To this purpose, resonant auxiliary oscillators (local resonators) can conveniently be introduced to realize acoustic metamaterials, featured by an enlarged configuration space of active degrees-of-freedom and a richer variety of tunable mechanical parameters [14],[25],[27],[28],[29]. Among the others, common customization criteria are the presence of selected harmonics in the band structure at a certain wavenumber [30],[31], the opening of maximum-amplitude band gaps in the lowest possible frequency range [32],[33],[34],[35],[36], the maximal sensitivity of the spectrum to microstructural defects [37],[38], the occurrence of negative refraction properties [39],[40].

Asymptotic techniques may allow the multiparametric approximation of the direct and inverse dispersion problem for low-dimensional lagrangian models. Consequently, the conditions for the existence of pass and stop bands can be determined in a suited analytical – although approximate – form [38]. The relative optimization analyses may highlight how synthetic lagrangian models may possess a low-dimensional parameter space, insufficient for the search of a satisfying solution for inverse spectral problems. The present paper is devoted at exploring the dispersion properties of the tetrachiral material in the larger parameter space obtainable by removing some of the simplifying mechanical assumption limiting the simpler lagrangian model. Two alternative continuous models (high-fidelity first-order heterogeneous and equivalent homogenized micropolar continuum) are derived in parallel to the lagrangian beam-lattice formulation (Section 2). The frequency dispersion spectra resulting from all the models are compared to each other and cross-validated (Sections 3). The qualitative and qualitative agreement between asymptotic perturbation-based approximations and numerical spectral solutions is discussed (Paragraph 3.1). Parametric analyses concerning the effects of variations in the enlarged space of geometric and mechanical parameters on the acoustic and optical surfaces are carried out (Section 4). Consequently, a satisfying tuning of the micromechanical properties is employed to successfully verify the filtering functionality of the material in the forced wave propagation (Paragraph 4.1). Concluding remarks are finally pointed out.

## 2. Tetrachiral material

### 2.1 Beam lattice model

The class of chiral and antichiral cellular materials is characterized by a periodic tessellation of the bidimensional plane. The elementary cell is strongly characterized by a microstructure composed by stiff circular rings connected by flexible straight ligaments, arranged according to different planar geometries including the trichiral, hexachiral, tetrachiral, anti-trichiral, antitetrachiral topologies [20]. Among the others, the tetrachiral material is featured by a monoatomic centrosymmetric cell in which the central stiff and massive ring (or disk) is connected to four tangent flexible and light ligaments (figure 1a). The periodic square cell has side length $H$. Each ring is mechanically modeled as a rigid annular body with mass $M_r$, rotational inertia $J_r$, mean radius $R$ and tranversal width $t_r$, (figure 1b). Each ligament is mechanically modeled as a linear unshearable beam, with material density $\rho_b$, transversal width $t_b$ and natural length $L_b = H \cos\beta$, where the *chirality angle* $\beta = \arcsin(2R/H)$ is the ligament inclination angle with respect to the ideal line connecting the centers of adjacent rings. A linear elastic material, with Young's modulus $E_b$ is assumed for all beams.

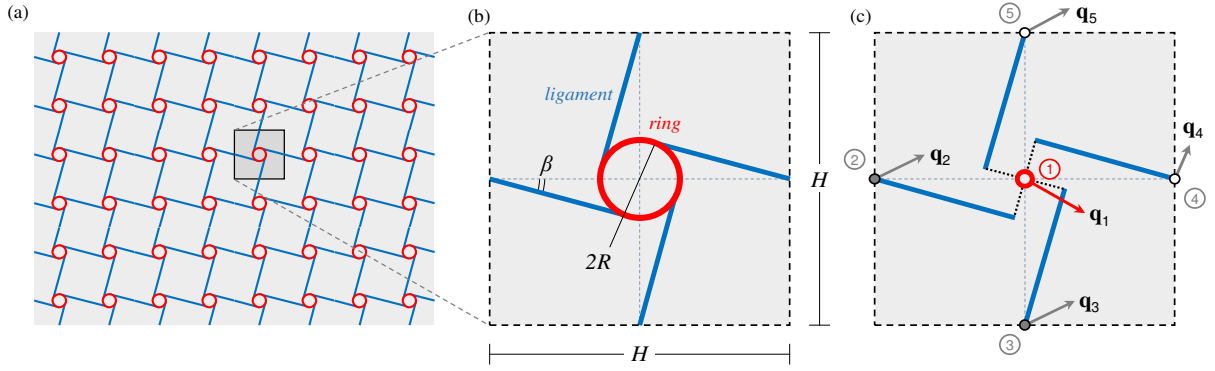

**Figure 1**. Tetrachiral metamaterial (a) repetitive planar pattern, (b) periodic cell, (c) beam lattice model.

The rigid body configuration is fully described by three planar *active* degrees-of-freedom, collected in the generalized displacement vector $\mathbf{q}_a = \mathbf{q}_1$ (figure 1c), referred to the internal located at the ring barycenter. Due to the geometric periodicity, the cell boundary crosses the midspan of all the four ligaments. Consequently four external nodes are located at the midpoint of all the cell sides, each one possessing three planar *passive* degrees-of-freedom collected in the displacement vector $\mathbf{q}_p = (\mathbf{q}_2, ..., \mathbf{q}_5)$.

Assuming the ligaments rigidly connected to the ring, a lagrangian beam lattice model can be formulated. The free undamped vibrations of the lagrangian model are governed by a linear equation, defined in the full configuration vector $\mathbf{q} = (\mathbf{q}_a, \mathbf{q}_p)$

$$\begin{bmatrix} \mathbf{M} & \mathbf{O} \\ \mathbf{O} & \mathbf{O} \end{bmatrix} \begin{pmatrix} \ddot{\mathbf{q}}_a \\ \ddot{\mathbf{q}}_p \end{pmatrix} + \begin{bmatrix} \mathbf{K}_{aa} & \mathbf{K}_{ap} \\ \mathbf{K}_{pa} & \mathbf{K}_{pp} \end{bmatrix} \begin{pmatrix} \mathbf{q}_a \\ \mathbf{q}_p \end{pmatrix} = \begin{pmatrix} \mathbf{0} \\ \mathbf{f}_p \end{pmatrix} \qquad (1)$$

where dot indicates differentiation with respect to time and $\mathbf{O}$ stands for different-size empty matrices. Adopting a lumped mass description, the non-null mass submatrix $\mathbf{M}$ is diagonal. The symmetric submatrices $\mathbf{K}_{aa}$ and $\mathbf{K}_{pp}$ describe the stiffness of the active and passive nodes, respectively. The rectangular submatrix $\mathbf{K}_{ap} = \mathbf{K}_{pa}^T$ account for the elastic coupling among the active and passive nodes. The mass and stiffness matrices are reported in details in the Appendix. The vector $\mathbf{f}_p$ collects the reactive forces exerted by the adjacent cells on the passive nodes. The passive displacement and force vectors can be ordered and partitioned as $\mathbf{q}_p = (\mathbf{q}_p^-, \mathbf{q}_p^+)$, $\mathbf{f}_p = (\mathbf{f}_p^-, \mathbf{f}_p^+)$ to separate the variables $(\mathbf{q}_p^-, \mathbf{f}_p^-)$, related to the left/bottom sides of the cell boundary (composed by the external nodes 2, 3 shown in figure 1c), from the variables $(\mathbf{q}_p^+, \mathbf{f}_p^+)$ related to the right/top sides (composed by the external nodes 4, 5). According to this decomposition, the dynamic (upper) part of the equation (1) can be written

$$\mathbf{M}\ddot{\mathbf{q}}_a + \mathbf{K}_{aa}\mathbf{q}_a + \mathbf{K}_{ap}^+\mathbf{q}_p^+ + \mathbf{K}_{ap}^-\mathbf{q}_p^- = \mathbf{0} \tag{2}$$

whereas the quasi-static (lower) part can be written

$$\begin{bmatrix} \mathbf{K}_{pa}^- \\ \mathbf{K}_{pa}^+ \end{bmatrix} \mathbf{q}_a + \begin{bmatrix} \mathbf{K}_{pp}^= & \mathbf{K}_{pp}^\mp \\ \mathbf{K}_{pp}^\pm & \mathbf{K}_{pp}^\# \end{bmatrix} \begin{pmatrix} \mathbf{q}_p^- \\ \mathbf{q}_p^+ \end{pmatrix} = \begin{pmatrix} \mathbf{f}_p^- \\ \mathbf{f}_p^+ \end{pmatrix} \tag{3}$$

According to the Floquet-Block theory for two dimensional discrete model [41], the quasi-periodic conditions governing the propagation of planar wave can be imposed on the passive displacement/forces at the cell boundary, requiring

$$\mathbf{q}_p^+ = \mathbf{L}_k \mathbf{q}_p^-, \quad \mathbf{f}_p^+ = \mathbf{L}_k \mathbf{f}_p^- \tag{4}$$

where $\mathbf{L}_k$ is a square *transfer* matrix that can be expressed in the diagonal block form

$$\mathbf{L}_k = \text{diag}\left(\exp(i k_1 H)\mathbf{I}, \exp(i k_2 H)\mathbf{I}\right) \tag{5}$$

where $k_1$ and $k_2$ are the two components of the wavevector $\mathbf{k} = (k_1, k_2)$, that is, the wavenumbers of the horizontally and vertically propagating waves, respectively.

The conditions (4) can be introduced in the quasi-static equation (3) to reduce the number of independent passive displacements [23]. Therefore, the linear quasi-static law

$$\mathbf{q}_p^- = \mathbf{R}_k \left(\mathbf{K}_{pa}^+ + \mathbf{L}_k \mathbf{K}_{pa}^-\right) \mathbf{q}_a, \quad \mathbf{f}_p^- = \mathbf{K}_{pa}^- + \left(\mathbf{K}_{pp}^= \mathbf{K}_{pa}^\mp \mathbf{L}_k\right) \mathbf{R}_k \left(\mathbf{K}_{pa}^+ + \mathbf{L}_k \mathbf{K}_{pa}^-\right) \mathbf{q}_a \tag{6}$$

govern the relations between the passive displacements or forces and the active degrees-of-freedom. The auxiliary block diagonal matrix $\mathbf{R}_k = -\left(\mathbf{L}_k \mathbf{K}_{pp}^\mp \mathbf{L}_k + \mathbf{L}_k \mathbf{K}_{pp}^= + \mathbf{K}_{pp}^\# \mathbf{L}_k + \mathbf{K}_{pp}^\pm\right)^{-1}$ is determined by the inversion of the non-singular sum between brackets.

Forcing the quasi-static relations (6) into the equation (2), the wave propagation through the material is fully governed in the configuration space of the active displacements by the equation of motion

$$\mathbf{M}\ddot{\mathbf{q}}_a + \mathbf{K}\mathbf{q}_a = \mathbf{0} \tag{7}$$

where the $\mathbf{k}$-dependent Hermitian matrix $\mathbf{K} = \mathbf{K}_{aa} + \left(\mathbf{K}_{ap}^+ \mathbf{L}_k + \mathbf{K}_{ap}^-\right) \mathbf{R}_k \left(\mathbf{K}_{pa}^+ + \mathbf{L}_k \mathbf{K}_{pa}^-\right)$.

Denoting $\Omega$ the unknown circular frequency, the oscillatory solution $\mathbf{q}_a = \boldsymbol{\psi}_a \exp(i\Omega t)$ can be imposed. Therefore, eliminating the dependence on time, the in-plane wave propagation is governed by the linear eigenproblem

$$(\mathbf{K} - \Lambda \mathbf{M})\boldsymbol{\psi}_a = \mathbf{0} \tag{8}$$

in the unknown eigenvalues $\Lambda = \Omega^2$ and eigenvectors $\boldsymbol{\psi}_a$. The eigensolution is composed by three eigenpairs, each made of a real-valued eigenvalue $\Lambda_h$ and a complex-valued eigenvector $\boldsymbol{\psi}_h$ $(h = 1,...,3)$. The eigenproblem (8) can be reformulated in the standard form

$$(\mathbf{H} - \Lambda \mathbf{I})\boldsymbol{\varphi}_a = \mathbf{0} \tag{9}$$

where the matrix $\mathbf{H} = \mathbf{Q}^{-T}\mathbf{K}\mathbf{Q}^{-1}$ and the auxiliary eigenvector $\boldsymbol{\varphi}_a = \mathbf{Q}\boldsymbol{\psi}_a$ are obtained by decomposing the mass matrix in the form $\mathbf{M} = \mathbf{Q}^T\mathbf{Q}$. The beam lattice model will be also referred to as *coarse* model in the following.

*2.2 Microscopic first order continuum model*

As alternative to the beam lattice model, the ring and the ligaments can be modeled at the microscopic scale in the framework of solid mechanics. Both the ring and the ligament bodies are modeled as first order continuum subject to Cauchy stresses induced by body forces $\mathbf{b}(\mathbf{x})$. A planar stress state is considered. Each material point is characterized by the displacement field $\mathbf{u}(\mathbf{x},t)$ and the partial differential equation governing the dynamic balance of a material point is

$$\nabla \cdot \boldsymbol{\sigma}(\mathbf{x}) + \mathbf{b}(\mathbf{x}) = \rho \ddot{\mathbf{u}}(\mathbf{x},t) \tag{10}$$

where $\rho$ is the mass density, $\ddot{\mathbf{u}}(\mathbf{x},t)$ is the acceleration of the material point and the stress tensor $\boldsymbol{\sigma}(\mathbf{x})$ can be related to the strain tensor $\boldsymbol{\varepsilon}(\mathbf{x})$ through the constitutive equation

$$\boldsymbol{\sigma}(\mathbf{x}) = \mathbb{C}^m(\mathbf{x})\boldsymbol{\varepsilon}(\mathbf{x}) \tag{11}$$

where $\boldsymbol{\varepsilon}(\mathbf{x}) = \text{sym}\nabla\mathbf{u}(\mathbf{x},t) = \frac{1}{2}\left[\nabla\mathbf{u}(\mathbf{x},t) + \nabla^T\mathbf{u}(\mathbf{x},t)\right]$ and $\mathbb{C}^m(\mathbf{x})$ is the fourth order elasticity tensor. Consistently with the beam lattice model, an elastic isotropic material is assumed. By exploiting the constitutive equation (11), the dynamic equation (10) becomes

$$\nabla \cdot (\mathbb{C}^m(\mathbf{x})\,\text{sym}\,\nabla\mathbf{u}(\mathbf{x},t)) + \mathbf{b}(\mathbf{x}) = \rho \ddot{\mathbf{u}}(\mathbf{x},t) \tag{12}$$

By applying the Fourier transform with respect to the time variable $t$, that is, $\mathcal{F}[\mathbf{u}(\mathbf{x},t)] = \int_{-\infty}^{+\infty} \mathbf{u}(\mathbf{x},t)e^{i\Omega t}dt = \hat{\mathbf{u}}(\mathbf{x})$ (with $\Omega$ angular frequency and $i^2 = -1$), the governing equation in the transformed space (Christoffel equation) is

$$\nabla \cdot (\mathbb{C}^m(\mathbf{x}) \operatorname{sym} \nabla \hat{\mathbf{u}}(\mathbf{x})) + \Omega^2 \rho \hat{\mathbf{u}}(\mathbf{x}) = \mathbf{0} \tag{13}$$

which is formally identical to the equation of motion (12) if a time harmonic dependence $\mathbf{u}(\mathbf{x},t) = \hat{\mathbf{u}}(\mathbf{x})e^{-i\Omega t}$ is assumed. According to the Floquet-Block theory for a two dimensional continuous model, the quasi-periodic conditions

$$\hat{\mathbf{u}}^+ = e^{i\mathbf{k}\cdot H}\hat{\mathbf{u}}^-, \qquad \hat{\boldsymbol{\sigma}}^+\mathbf{n}^+ = -e^{i\mathbf{k}\cdot H}\hat{\boldsymbol{\sigma}}^-\mathbf{n}^- \tag{14}$$

where $\mathbf{k}$ is the wavevector and $\mathbf{n}$ is the outward normal unit vector ($\mathbf{n}^+$ and $\mathbf{n}^-$ are defined, analogously to beam lattice model, on the right/top and on the left/bottom edges, respectively). The Floquet–Bloch problem is solved numerically via a Finite Element model [42]. Two dimensional triangular elements (with quadratic Lagrangian interpolation functions) are adopted. A direct solver based on the LU decomposition method is used. The microscopic first order continuum model will be also referred to as *fine* model in the following.

## 2.3 Homogenized micropolar continuum model

The continuous displacement field of a micropolar continuum model is $\mathbf{v}(\mathbf{x},t)$ and $\theta(\mathbf{x},t)$ that represent, respectively, the macro-displacement and the micropolar rotation of the reference cell located at $\mathbf{x}$ at the time $t$.

The equations of motion in the compact form are

$$\begin{aligned}
\nabla \cdot \mathbf{T} &= \rho \ddot{\mathbf{v}} \\
\nabla \cdot \mathbf{m} - \in_{3jh} \left(\mathbf{e}_j \otimes \mathbf{e}_h\right) : \mathbf{T} &= \eta \ddot{\theta}, \qquad j,h=1,2
\end{aligned} \tag{15}$$

where $\rho$ is the mass density of the equivalent homogeneous continuum, $\eta$ is the density of rotational inertia, $\mathbf{T}$ is the asymmetric macro-stress tensor, $\mathbf{m}$ is the couple-stress vector of the equivalent continuum and $\in_{3jh}$ is the Levi Civita symbol.

By introducing the curvature $\boldsymbol{\chi} = \nabla \theta$ and the micropolar asymmetric strain tensor $\boldsymbol{\Gamma} = \nabla \mathbf{v} - \mathbf{W}(\theta)$, where $\nabla \mathbf{v}$ is the displacement gradient and the macro-rotation micropolar tensor $\mathbf{W}$ is defined as $\mathbf{W} = w_{jh}\mathbf{e}_j \otimes \mathbf{e}_h = -\in_{3jh} \theta \mathbf{e}_j \otimes \mathbf{e}_h$, the constitutive relations of micropolar continuum are

$$\begin{aligned}
\mathbf{T} &= \mathbb{E}_s \boldsymbol{\Gamma} + \mathbb{Y}_s \boldsymbol{\chi} \\
\mathbf{m} &= \mathbb{Y}_s^T \boldsymbol{\Gamma} + \mathbf{E}_s \boldsymbol{\chi}
\end{aligned} \tag{16}$$

where $\mathbb{E}_s$, $\mathbb{Y}_s$ and $\mathbf{E}_s$ are, respectively, the fourth, third and second order elasticity tensors of the equivalent homogeneous continuum.

By exploiting the constitutive equations (16) the equations motion result as

$$\nabla \cdot (\mathbb{E}_s \Gamma + \mathbb{Y}_s \chi) = \rho \ddot{\mathbf{v}}$$
$$\nabla \cdot (\mathbb{Y}_s^T \Gamma + \mathbb{E}_s \chi) - \in_{3jh} (\mathbf{e}_j \otimes \mathbf{e}_h) : (\mathbb{E}_s \Gamma + \mathbb{Y}_s \chi) = \eta \ddot{\theta}, \qquad j, h = 1, 2 \qquad (17)$$

In case of centrosymmetric lattice, the third order elasticity tensor $\mathbb{Y}_s$ is identically null and the equations (17) can be written as

$$\left[ E_{ijhk} \left( v_{h,k} + \in_{3hk} \theta \right) \right]_{,j} = \rho \ddot{v}_i$$
$$\left( E_{ij} \theta_{,j} \right)_{,i} - \in_{3jh} E_{jhrs} \left( v_{r,s} + \in_{3rs} \theta \right) = \eta \ddot{\theta}, \qquad i, j, h, k, r, s = 1, 2 \qquad (18)$$

where the elastic constants $E_{ijhk}$ and $E_{ij}$ can be obtained through a high frequency dynamic homogenization similar to that proposed in [23].

By applying to (17) the Fourier transform respect to the time variable $t$ or by considering the harmonic motion $\mathbf{v} = \hat{\mathbf{v}} \exp[i(\mathbf{kx} - \Omega t)]$ and $\theta = \hat{\theta} \exp[i(\mathbf{kx} - \Omega t)]$, where the polarization vector is defined by the collection of $\hat{\mathbf{v}}$ and $\hat{\theta}$, the governing equation in the transformed space (Christoffel equation) is

$$\mathbf{C}_{Hom}(\mathbf{k}, \Omega) \hat{\mathbf{V}} = \mathbf{0} \qquad (19)$$

where $\hat{\mathbf{V}} = [\hat{\mathbf{v}}^T, \hat{\theta}]$. In case of centrosymmetric lattice, the equation (19) can be written as

$$E_{ijhk} \left( k_j k_k \hat{v}_h - \in_{3hk} i k_j \hat{\theta} \right) - \rho \Omega^2 \hat{v}_i = 0$$
$$E_{ij} k_i k_j \hat{\theta} + \in_{3jh} E_{jhrs} \left( i k_s \hat{v}_r + \in_{3rs} \hat{\theta} \right) - \eta \Omega^2 \hat{\theta} = 0, \qquad i, j, h, k, r, s = 1, 2 \qquad (20)$$

Can be demonstrated that the second order expansion of the matrix of the micropolar model $\mathbf{C}_{Hom}(\mathbf{k}, \Omega)$ in the wave vector $\mathbf{k}$ corresponds to an approximation the corresponding one from the Lagrangian system $\mathbf{C}_{Lag}(\mathbf{k}, \Omega)$ [24]

$$\mathbf{C}_{Lag}(\mathbf{k}, \Omega) = A_{cell} \mathbf{C}_{Hom}(\mathbf{k}, \Omega) + O\left(|\mathbf{k}|^3\right) \qquad (21)$$

where, considering the eigenvalue problem of the beam lattice model (8), the matrix of the Lagrangian system is $\mathbf{C}_{Lag}(\mathbf{k}, \Omega) = \mathbf{K} - \Lambda \mathbf{M}$ and $A_{cell}$ is the area of the periodic cell. The homogenized micropolar continuum model will be also referred to as *homogenized* model in the following.

## 3. Wave propagation

*3.1 Dispersion spectrum*

The $\mathbf{k}$-dependent solution of the eigenproblem (8) gives the dispersion relations $\Omega_h(\mathbf{k})$ characterizing the coarse model ($h=1…3$). Alternatively, $\mathbf{k}$-dependent solution of the Cristoffel equation (13) coupled with the conditions (14) gives the dispersion relations $\Omega_h(\mathbf{k})$ characterizing the fine model ($h \in \mathbb{N}$). The dispersion surfaces are obtained letting the wavevector $\mathbf{k}$ vary in the entire first Brillouin zone [41], corresponding to the square domain $\mathscr{D}^H = [-\pi/H, \pi/H] \times [-\pi/H, \pi/H]$. The dispersion functions $\Omega_h(\Xi)$ can be defined by introducing a curvilinear abscissa $\Xi$ spanning the closed boundary $\partial \mathscr{B}_1^H$ of the triangular subdomain $\mathscr{B}_1^H$. The corresponding curves obtained under variation of the abscissa $\Xi$ over the entire range $[0, (2\pi + \sqrt{2}\pi)/H]$ fully characterize the dispersion spectrum of one or the other models. The numerical solution for the fine model has been based on a sufficiently fine discretization of the finite element model, selected after a proper convergence analysis. Moreover, the triangular boundary $\partial \mathscr{B}_1^H$ has been divided in one hundred equidistant $\Xi$-points to obtain a sufficient resolution in the dispersion curves.

Furthermore, the exact solution of the eigenproblem (8) has been approximated with the solution of the Cristoffel equation (19) related to the homogenized model. In particular it can be demonstrated that the dispersion curves of homogenized model coincide with those of the coarse model, if a second-order Taylor $\mathbf{k}$-expansion (centered in $\mathbf{k}=\mathbf{0}$) of the $\mathbf{K}$-matrix is performed [24],[43].

Introducing the nondimensional wave vector $\boldsymbol{\beta} = (\beta_1, \beta_2)$ where $\beta_1 = k_1 H$ and $\beta_2 = k_2 H$, a comparison between the dispersion spectra of the fine and coarse models (figure 2) and between the fine and homogenized models (figure 3a) is shown. The comparison is carried out in nondimensional form by introducing the dispersion functions $\omega_h(\xi)$ relating the nondimensional variables

$$\omega = \frac{\Omega}{\Omega_c} , \quad \xi = \Xi H , \tag{22}$$

where $\Omega_c^2 = E_b (\rho_r H^2)^{-1}$ is a reference frequency that can be set to be unitary without loss of generality. The independent variable $\xi$ varies in the range $[0, 2\pi + \sqrt{2}\pi]$ and represents the curvilinear abscissa spanning the boundary $\partial \mathscr{B}_1$ of the triangular subdomain $\mathscr{B}_1$ of the nondimensional domain $\mathcal{D} = [-\pi,\pi] \times [-\pi,\pi]$ of the first Brillouin zone.

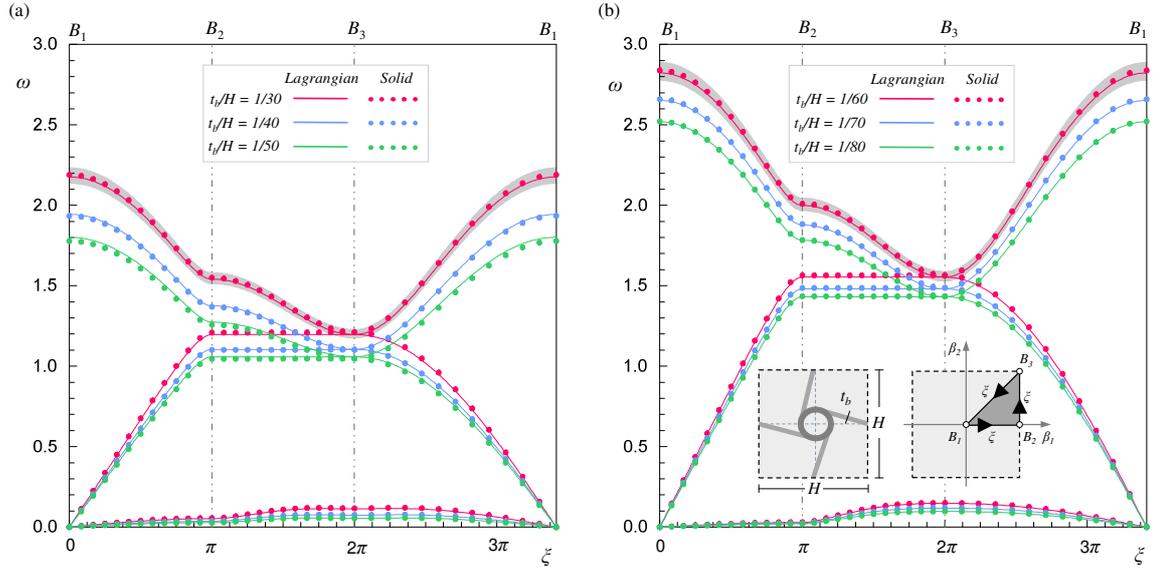

**Figure 2.** Comparison between the dispersion spectra of the beam lattice and the microstructural first order models for the tetrachiral material (with $t_r/t_b = 5$) for different values of the ligament slenderness $t_b/H$: (a) $R/H = 1/5$, (b) $R/H = 1/10$.

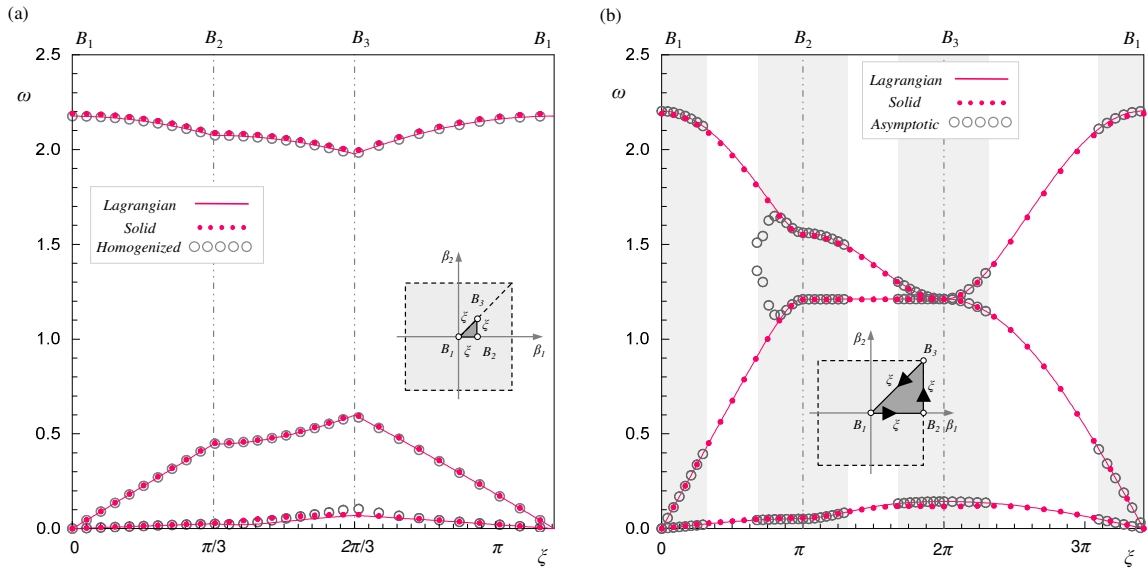

**Figure 3.** Dispersion spectra for the tetrachiral material ($t_b/H = 1/30$, $t_r/t_b = 5$, $R/H = 1/5$): (a) Comparison among the coarse (Lagrangian), fine (Solid) and homogenized models, (b) Comparison between the fine (Solid) model and the exact and asymptotically approximate solutions for the coarse (Lagrangian) model.

The good matching covers all the boundary $\partial \mathcal{B}_1$ in the comparison between the coarse and the fine models for different values of the nondimensional parameter $t_b/H$, accounting for the ligament slenderness (figure 2). In particular, a good agreement is obtained in the gray region around the red dispersion curve with the highest-frequency, corresponding to different effective lengths of the ligaments $L_e = \alpha L_b$ in the beam lattice model. The gray region is the envelope of the dispersion curves obtained for the multiplier $\alpha$ varying in the range $[0.78, 0.85]$ for the figure 2a and in the range $[0.88, 0.95]$ for the figure 2b. Differently, the homogenized model is found to well-approximate the dispersion spectrum of the fine model along the boundary of a subdomain of the triangular $\mathcal{B}_1$ zone (figure 2b).

*3.2 Perturbation solution*

Within the framework of beam lattice models, the step-by-step construction of the dispersion surfaces $\Omega_h(\mathbf{k})$ rapidly tends to demand excessive computational resources as the model dimension increases and the $\partial \mathcal{B}_1^H$-discretization becomes finer. Therefore, perturbation methods can represent an efficient alternative to the numerical solution of the eigenproblem governing the wave dispersion. Each eigenvalue $\Lambda_h(\mathbf{k}) = \Omega_h(\mathbf{k})$ satisfying the eigenproblem (9) can be regarded as one of the zeroes (with multiplicity $m_h$) of the characteristic function $Q(\Lambda, \mathbf{k}) = \det(\mathbf{H}(\mathbf{k}) - \Lambda \mathbf{I})$ in the domain of positive real $\Lambda$-values. Therefore, fixed the set $\mathbf{p}^*$ of the microstructural parameters describing the periodic cell, each dispersion surface can be determined by individually following a certain zero of the characteristic function, under variation of the wavevector $\mathbf{k}$ in the square Brillouin domain.

Within this mathematical context, perturbation techniques furnish analytical – although asymptotically approximate – expressions of the dispersion functions. Naturally, the availability of explicit functions $\Lambda_h(\mathbf{k})$ may reduce the algorithmic effort required by numerical continuation traditionally applied to the solutions of the characteristic equation $Q(\Lambda, \mathbf{k}) = 0$. It its basic form, a perturbation technique starts with the selection of a reference (unperturbed) wavevector $\mathbf{k}^*$ in the $\mathcal{B}_1^H$-domain. The corresponding reference eigenvalues $\Lambda^*$ of the matrix $\mathbf{H}^* = \mathbf{H}(\mathbf{k}^*)$ are assumed exactly (analytically or numerically) known. Even if not mandatory, a convenient selection of the reference wavevector $\mathbf{k}^*$ could be preferable to enhance the effectiveness and the validity of the asymptotic approximation (see also [44],[45]). For instance, the typical choice $\mathbf{k}^* = \mathbf{0}$ often allows the analytical assessment of all the reference eigenvalues $\Lambda^*$ satisfying the characteristic equation $Q(\Lambda, \mathbf{k}^*) = 0$.

In the most general case, perturbation techniques can readily treat multi-variable perturbations. Indeed, small-amplitude two-parameter perturbations $\varepsilon \mathbf{k}' = \mathbf{k} - \mathbf{k}^*$ can be considered to span all the directions of the bi-dimensional $\mathscr{B}_1^H$-domain in the neighborhood of $\mathbf{k}^*$ (since $\varepsilon \ll 1$ represents an auxiliary nondimensional parameter regulating the perturbation smallness). If necessary, higher-dimension perturbation vectors can also be adopted to asymptotically explore enlarged parametric spaces [23].

To the specific purposes of the present work, exploring the mono-dimensional boundary $\partial \mathscr{B}_1$, spanned by the nondimensional abscissa $\xi$, may be sufficient. Consequently, the unperturbed reference point in the boundary $\partial \mathscr{B}_1$ can be identified by the particular abscissa $\xi^*$, and the $\xi^*$-neighborhood can be explored by the local coordinate $z = \xi - \xi^*$. Thus – within the limits of the boundary $\partial \mathscr{B}_1$ – the change-of-variable $\xi = \xi^* + z$ allows to express the nondimensional characteristic function in the form $F(\lambda, z)$, where the $z$-variable acts as single perturbation parameter (under the assumption $z \ll 1$) and $\lambda = \omega^2$ is the nondimensional eigenvalue.

Under the assumption of sufficient regularity of the dispersion functions, each exact eigenvalue can tentatively be approximated by a series function $\lambda(z)$ of integer $z$-powers

$$\lambda(z) = \lambda^* + \sum_n \lambda^{(n)} z^{(n)} = \lambda^* + \dot{\lambda} z + \ddot{\lambda} z^2 + ... + \lambda^{(n)} z^{(n)} + ... \qquad (23)$$

where the coefficient $\lambda^{(n)}$ (multiplied by the factorial $n!$) are known as the *eigensensitivities* (with respect to the perturbation $z$) of the eigenvalues, and can also be regarded as the *unknown* $n$-th $z$-derivative (evaluated at $z = 0$) of the exact but implicit eigenvalue function $F(\lambda, z) = 0$.

Once the series $\lambda(z)$ has been established, the characteristic function becomes a composite single-variable function $G(z) = F(\lambda(z), z)$, which admits the Taylor expansion in $z$-powers

$$G(z) = G^* + \sum_n \frac{G^{(n)}}{n!} z^{(n)} = G^* + \dot{G} z + \frac{\ddot{G}}{2} z^2 + ... + \frac{G^{(n)}}{n!} z^{(n)} + ... \qquad (24)$$

where $G^* = F(\lambda^*, 0)$ is certainly null, as far as $\lambda^*$ is known through the relation $\lambda^* = \Lambda^* / \Omega_c^2$ where $\Lambda^*$ belongs to the $\mathbf{H}^*$-eigenspectrum by hypothesis. The generic higher-order coefficient $G^{(n)}$ can be recognized as the *n*-th $z$-derivative (evaluated at $z=0$) of the function $G(z)$. In general, these derivatives require the recursive application of the chain rule for the differentiation of single-variable composite functions.

Each series coefficient $G^{(n)}$ is a complete *n*-degree polynomial of all the unknown coefficients of the eigenvalue expansion (19) up to $\lambda^{(n)}$. The $G^{(n)}$-coefficients multiplying the lowest $z$-powers are

$$z^1: \quad \dot{G} = \dot{\lambda} F^{(1,0)} + F^{(0,1)} \tag{25}$$

$$z^2: \quad \ddot{G} = 2\ddot{\lambda} F^{(1,0)} + (\dot{\lambda})^2 F^{(2,0)} + 2\dot{\lambda} F^{(1,1)} + F^{(0,2)} \tag{26}$$

where the synthetic notation $F^{(h,k)} = \partial_\lambda^h \partial_z^k F(\lambda, z)$ has been adopted for the partial derivatives of the characteristic function $F(\lambda, z)$, and evaluation at $(\lambda = \lambda^*, z = 0)$ is understood. A recursive form of the generic *n*-th coefficient, multiplying the $z^n$-power, can be found in [38].

The characteristic equation $G(z) = F(\lambda(z), z) = 0$ is asymptotically satisfied by zeroing each $z^n$-order coefficient $G^{(n)}$. Thus, a chain of *n* ordered equations (*perturbation equations*) is generated, where the eigensensitivities $\lambda^{(n)}$ are the unknowns to be determined. Starting with the zeroth-order solution, given by the known eigenvalues $\lambda^*$ (*generating solution*), each perturbation equation of the chain involves a single unknown, that is, one of the higher-order coefficients. Depending on the algebraic multiplicity $m^*$ of the generic $\lambda^*$-eigenvalue, two fundamental cases occur (see also [45])

- Simple eigenvalue: if $\lambda^*$ is a simple root (algebraic multiplicity $m^* = 1$) for the equation $G(0) = F(\lambda^*, 0) = 0$, then the coefficient $F^{(1,0)} \neq 0$. Hence, the $z^1$-order equation (24) is linear in the unknown $\dot{\lambda}$, the $z^2$-order equation (25) is linear in the unknown $\ddot{\lambda}$, and so on. Therefore, the cascade solutions (null if the numerator vanishes) for the lowest order equations are

$$z^1: \quad \dot{\lambda} = -\frac{F^{(0,1)}}{F^{(1,0)}} \tag{27}$$

$$z^2: \quad \ddot{\lambda} = -\frac{(\dot{\lambda})^2 F^{(2,0)} + 2\dot{\lambda} F^{(1,1)} + F^{(0,2)}}{2 F^{(1,0)}} \tag{28}$$

and, by extension, the $z^n$-order equation allows the determination of the *n*-th coefficient $\lambda^{(n)}$. A recursive form of the *n*-th coefficient $\lambda^{(n)}$ can be found in [38].

- *Double (semi-simple) eigenvalue*: if $\lambda^*$ is a double root (algebraic and geometric multiplicity $m^* = 2$) for the equation $G(0) = F(\lambda^*, 0) = 0$, then the coefficient $F^{(1,0)} = 0$, but $F^{(2,0)} \neq 0$. Since $\lambda^*$ must be non-defective (semi-simple), it can be proved that $F^{(0,1)} = 0$. Consequently, the $z^1$-order equation (24) is trivially satisfied, but leaves $\dot{\lambda}$ undetermined. Such an indetermination is cleared by the $z^2$ - order equation (25), which is a quadratic in the $\dot{\lambda}$-unknown only, since the null multiplier $F^{(1,0)}$ affects the other unknown $\ddot{\lambda}$. Thus, the lowest order equations give

$$z^1: \quad \dot{\lambda} \text{ is undetermined} \tag{29}$$

$$z^2: \quad \dot{\lambda}_\pm = -\frac{F^{(1,1)} \pm \sqrt{(F^{(1,1)})^2 - F^{(2,0)} F^{(0,2)}}}{F^{(2,0)}} \tag{30}$$

where the pair $\dot{\lambda}_\pm$ splits the double root $\lambda^*$ in two distinct eigenvalues $\lambda^* + z\dot{\lambda}_\pm + O(z^2)$. If the radical vanishes, in consequence of the particular sub-case $(F^{(1,1)})^2 = F^{(2,0)} F^{(0,2)}$, the splitting of the double root is postponed to higher-orders. In the general case, the higher unknowns $\lambda_\pm^{(n)}$ are determined by linear $z^{n+1}$-order equations, solved for one or the other of the $\dot{\lambda}_\pm$-values.

When the perturbation technique is applied to approximate the band structure of the tetrachiral material, the three vertices $B_1$ (corresponding to $\xi^* = 0$ or $\boldsymbol{\beta}^* = \mathbf{k}^* H = (0,0)$), $B_2$ (corresponding to $\xi^* = \pi$ or $\boldsymbol{\beta}^* = \mathbf{k}^* H = (\pi, 0)$) and $B_3$ (corresponding to $\xi^* = 2\pi$ or $\boldsymbol{\beta}^* = \mathbf{k}^* H = (\pi, \pi)$) of the triangular boundary $\partial \mathcal{B}_1$ have been employed as reference points to start the perturbation analysis. The fourth-order asymptotic approximation of the dispersion curves (determined by the eigensensitivies reported in [38]) are represented by the black circles in Figure 3. The comparison with the exact solutions obtained from the lagrangian model (red lines) and the solid model (red dots) shows a fine agreement over large extents of the boundary $\partial \mathcal{B}_1$, centered at the three reference points. Coherently with the intrinsic nature of the perturbation solutions, the approximation accuracy tends to decay with the distance from the reference point. However, a satisfying accuracy can be observed to persist up to $z$-values close or even greater than unity (the gray zones), that is, beyond the limits of the smallness assumption ($z << 1$).

## 4. Parametric analysis

The parametric analyzes are run for valuate the effects of the variations in the mechanic and geometric parameters on the acoustic and optic surfaces of the Floquet-Bloch spectrum. Given the good agreement obtained by the comparison of the dispersion spectrum of the different models, the analyzes are performed using the finite element solution of the fine Cauchy model that is easier to handle and suited to deal with a greater variety of cases. The tetrachiral material selected as reference for the parametric analyzes is characterized by empty ring and massless ligaments with same elastic properties.

Some preliminary analyzes are carried out by varying the independent nondimensional geometric parameters

$$\frac{t_b}{H}, \quad \frac{t_r}{t_b}, \quad \frac{R}{H} \tag{31}$$

that depend on the ring radius (associated to the chirality) and on the width of the ligaments and the ring. Increasing the width of the ligaments, the frequencies grow up in all investigation range and the distance between the optic and the acoustic surfaces increases creating a partial band gap with growing amplitude in the interval $0 \leq \xi \leq \pi$. Instead, increasing the radius or the width of the ring the frequencies decrease. When the width of the ring is small and therefore the ring is thin and flexible, an enrichment of the dispersion spectrum, in the low-frequency range, with new dispersion surfaces associated to ring-deforming waveforms can be observed. In this case, the hypothesis of rigid body used in the beam lattice model loses validity [46].

These analyzes highlight some issues in satisfying the conditions of the existence of a full band gap in the low-frequency range. These conditions are investigated numerically and determined in analytical form asymptotically as an inequality between the inertial characteristics of the ring and the slenderness of the ligaments [8],[25]. The physical realization of these conditions would require a technical arrangement like a material with functionally-graded elastic properties with tapered cross section or other modifications that destroy the invariant properties in the out-of-plane direction. It is possible to achieve a band gap at a target frequency using metamaterials with inertial resonators [38].

The possibility of obtaining the same result is analyzed either by enlarging the parameter space or by removing some simplifying hypotheses of the reference model. The frequency corresponding to the vertex $B_2$ of the triangular boundary $\partial \mathscr{B}_1$, where the second acoustic surface and the first optical surface collide given rise to a point with a double frequency, is chosen as target center frequency. The independent nondimensional parameters considered for these parametric analyzes are

$$\frac{\rho_i}{\rho_r}, \quad \frac{\rho_b}{\rho_r}, \quad \frac{E_{b1}}{E_{b2}} \tag{32}$$

where the parameter $\rho_i/\rho_r$ represents the mass density ratio between an intra-ring filler and the ring material, $\rho_b/\rho_r$ is the mass density ratio between the ligament and the ring material and the parameter $E_{b1}/E_{b2}$ is the ratio between the Young's Moduli of two ligaments made of different materials. The analysis cases and the parameter range variations are summarized in table 1.

Table 1: Fixed and free parametric values in the cases of the parametric analysis

|  | $t_b/H$ | $t_r/t_b$ | $R/H$ | $\rho_i/\rho_r$ | $\rho_b/\rho_r$ | $E_{b1}/E_{b2}$ |
|---|---|---|---|---|---|---|
| Reference Case | [1/100, 1/10] | [1, 3] | [1/10, 1/3] | 0 | 0 | 1 |
| Case 1 | 1/20 | 3 | 1/3 | 0 | [1/10, 30] | 1 |
| Case 2 | 1/20 | 3 | 1/3 | [1, 30] | [1/10, 30] | 1 |
| Case 3 | 1/20 | 3 | 1/3 | [1, 10] | 0 | [1, 30] |

The dispersion spectrum of the reference case for parameter values of $t_b/H = 1/20$, $t_r/t_b = 3$, $R/H = 1/3$ is shown in figure 4. There is no band gap at the low-frequency range and the double frequency, used as target center frequency for the follow analyzes, is 1.3. For the subsequent cases analyzed, the values of the parameters (31) are fixed as in the example shown in figure 4.

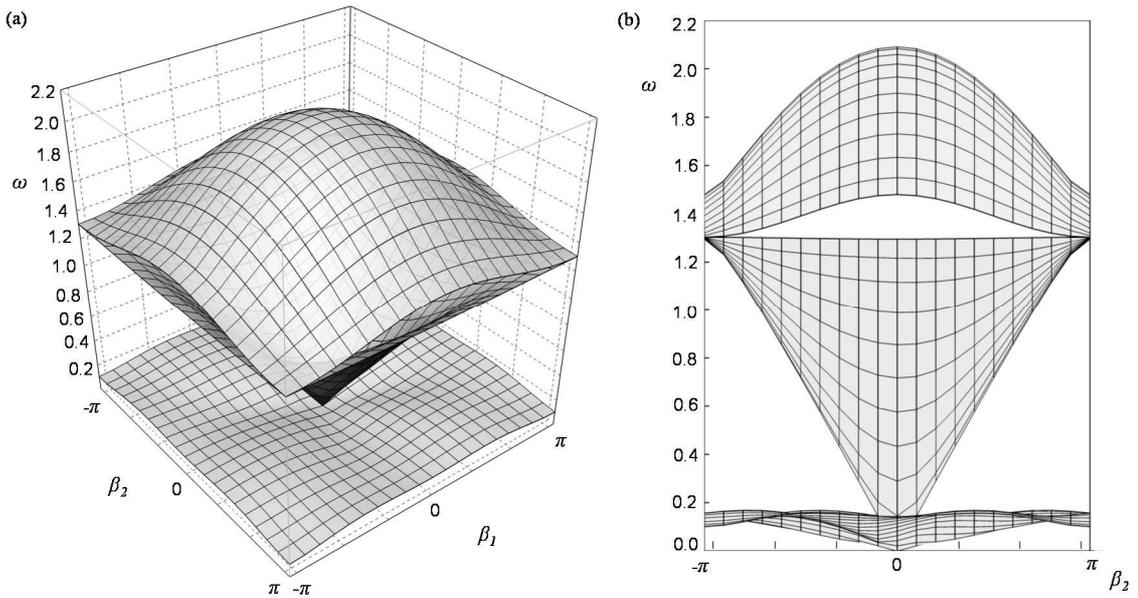

**Figure 4**. Dispersion spectrum with the parameters $t_b/H = 1/20$, $t_r/t_b = 3$, $R/H = 1/3$.

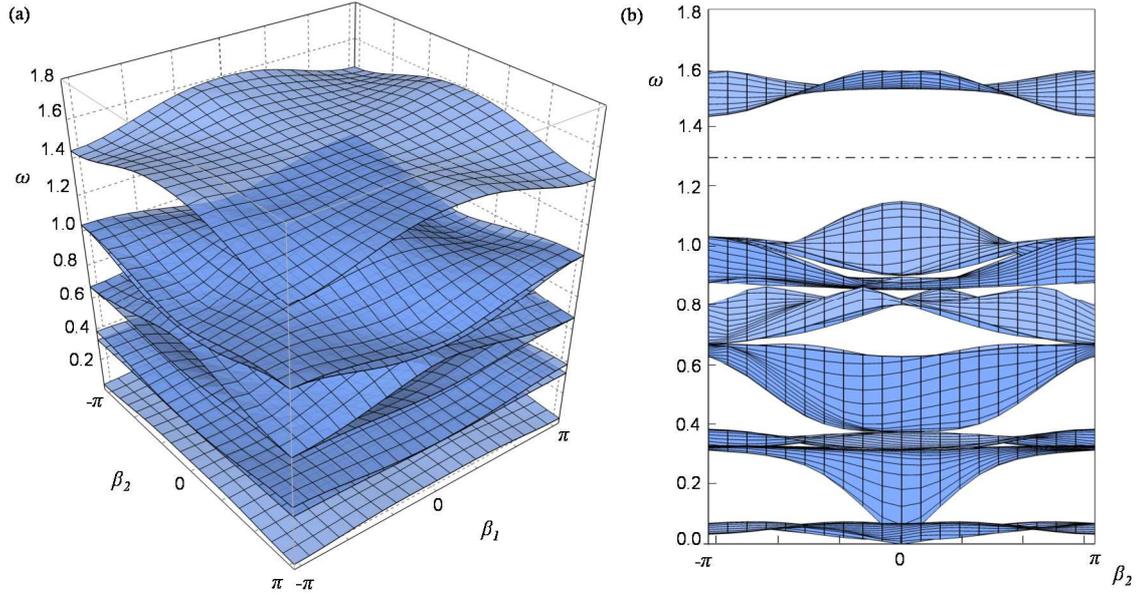

**Figure 5**. Dispersion spectrum of a tetrachiral material with massive ligaments and parameter $\rho_b/\rho_r = 18$.

In the case 1 the tetrachiral material has the massive ligaments and an empty ring. The parameters $\rho_i/\rho_r$ and $E_{b1}/E_{b2}$ are fixed and the ratio $\rho_b/\rho_r$ changes in the range reported in table 1. Increasing the mass density of the ligaments, the frequencies decrease and an enrichment of the dispersion spectrum with more dispersion optic surfaces in the low-frequency range is observed (figure 5a). It is possible to obtain, between the optic surfaces, a total band gap around the target center frequency. For $\rho_b/\rho_r = 18$ a band gap borns between the surfaces 7 and 8 (figure 5b).

In the case 2 the tetrachiral material has the massive ligaments and an intra-ring heavy filler. The parameter $E_{b1}/E_{b2}$ is fixed and the ratios $\rho_i/\rho_r$ and $\rho_b/\rho_r$ change in the ranges reported in table 1. Increasing the intra-ring filler mass density the frequencies decrease. Furthermore, increasing the mass density of the ligaments, the frequencies decrease much more and an enrichment of the dispersion spectrum with more dispersion optic surfaces in the low-frequency range is observed (figure 6a). It is possible to achieve, between the optic surfaces, a total band gap around the target center frequency. For $\rho_i/\rho_r = 10$ and $\rho_b/\rho_r = 16$ a band gap borns between the surfaces 7 and 8 (figure 6b). Another band gap is obtained between the surfaces 5 and 6 (figure 6b).

In the case 3 the tetrachiral material has an intra-ring heavy filler, two ligaments (green ligaments in the figure 7) with Young's Modulus equal to $E_{b1}$ and the other ligaments (grey ligaments in the figure 7) with Young's Modulus equal to $E_{b2}$. The ratio $\rho_b / \rho_r$ is fixed and the parameters $\rho_i/\rho_r$ and $E_{b1}/E_{b2}$ change in the ranges reported in table 1. Increasing the parameter $E_{b1}/E_{b2}$ the frequencies of the optic surface grow up and increasing the ratio $\rho_i/\rho_r$ a total band gap between the second acoustic surface and the first optic surface in the low-frequency range can be observed. For $\rho_i/\rho_r=10$ and $E_{b1}/E_{b2}=8$ a band gap between the surfaces 2 and 3 borns (figure 7).

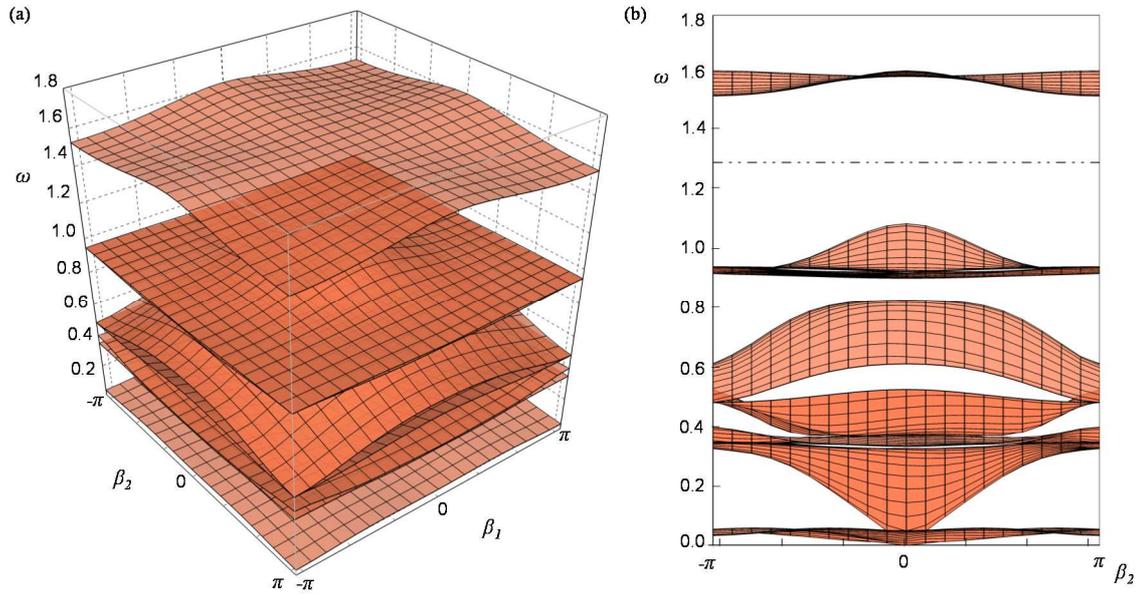

**Figure 6**. Dispersion spectrum of a tetrachiral material with massive ligaments, intra-ring heavy filler and parameters $\rho_b / \rho_r = 16$ and $\rho_i / \rho_r = 10$.

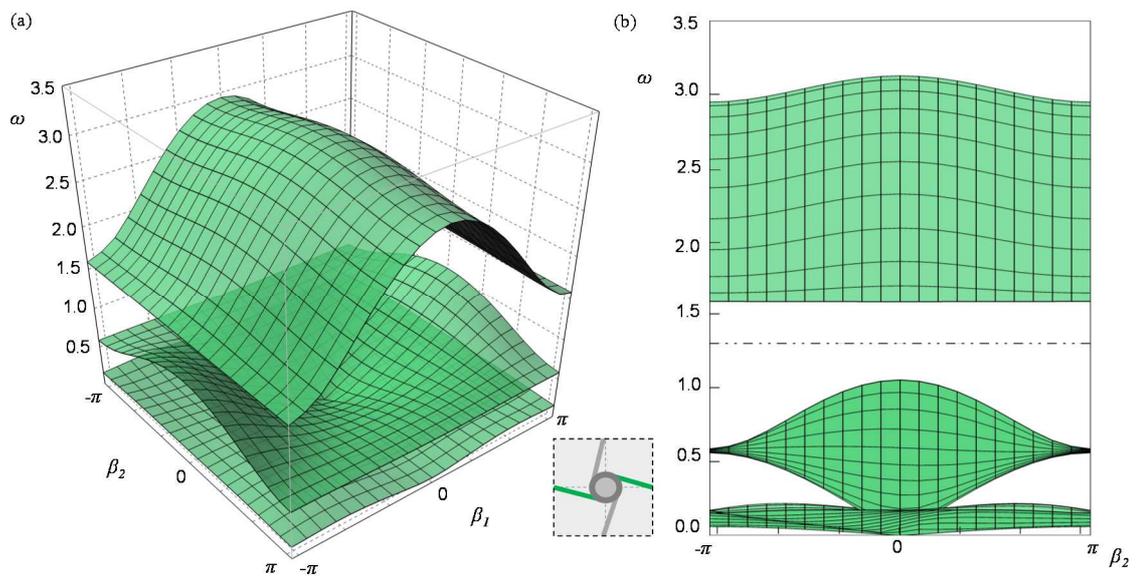

**Figure 7**. Dispersion spectrum of a tetrachiral material with intra-ring heavy filler, inhomogeneous massless ligaments and parameters $E_{b1} / E_{b2} = 8$ and $\rho_i / \rho_r = 10$.

*4.1 Performance as acoustic filters*

A material that stops the propagation of harmonic waves with frequencies that do not belong to the spectrum, therefore a material that has a dispersion spectrum in which there is a total band gap, behaves as an acoustic filter for elastic waves. In this regard, the variation of the adimensional parameters (32) causes changes on the band gap amplitude $\Delta$ and on the center frequency $\gamma$. Therefore, the controlled variation of these parameters can be considered a mechanical tuning of the acoustic filter realized by tetrachiral material.

In the investigated parameter region, decreasing the mass density ratio $\rho_b/\rho_r$ between the ligaments and the ring material, the band gap amplitude $\Delta$ increases for each intra-ring filler mass density. Furthermore, increasing the mass density ratio $\rho_i/\rho_r$ between an intra-ring filler and the ring material the band gap amplitude $\Delta$ grows up (figure 8a). A similar behavior can be observed for the center frequency in the figure 8b, where the pink band includes the values of the frequencies that are within the band gap for $\rho_i/\rho_r=10$. For instance, for the center frequency $\gamma = 1.3$ and $\rho_i / \rho_r = 10$, it is obtained $\rho_b / \rho_r \simeq 16$ (red dot in the figure 8b) and a band gap amplitude $\Delta \simeq 0.45$ (red dot in the figure 8a). For the same center frequency, other solutions with different band gap amplitudes can be achieved (green, blue and orange dots in the figure 8).

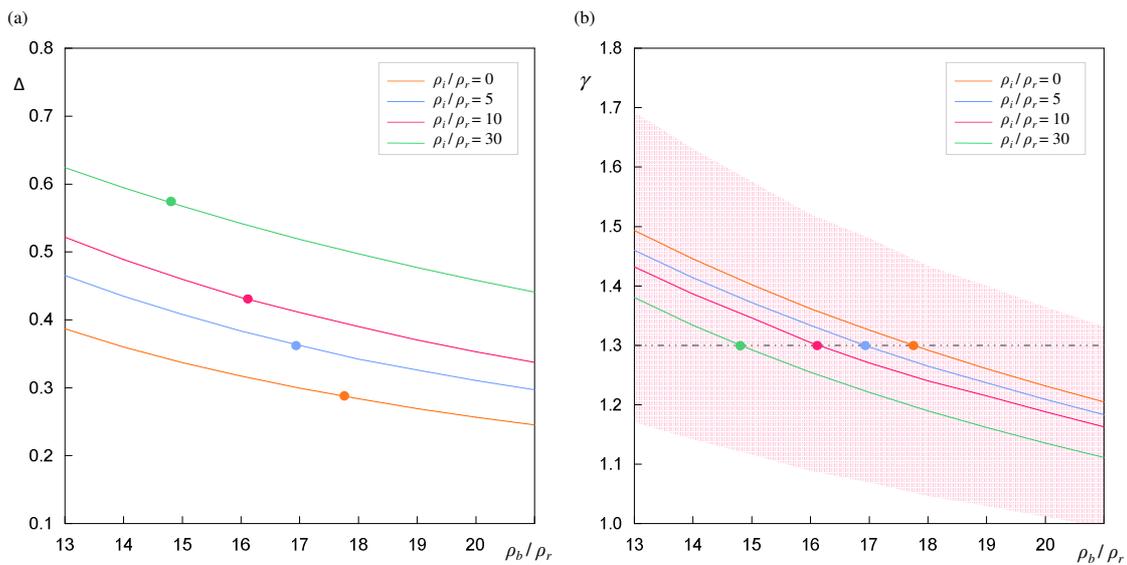

**Figure 8**. (a) Band gap amplitude, (b) Center frequency for a tetrachiral material with massive ligaments and intra-ring filler.

Alternatively, the band gap amplitude changes by tuning the ratio $E_{b1} / E_{b2}$ and by varying the intra-ring elastic filler mass density. Increasing the ratio $E_{b1} / E_{b2}$ the band gap amplitude grows up for each intra-ring filler mass density (figure 9a). In reverse, increasing the mass density ratio $\rho_i / \rho_r$ between an intra-ring filler and the ring material the band gap amplitude $\Delta$ decreases. For instance, for the center frequency $\gamma = 1.3$ and $\rho_i / \rho_r = 10$, it is obtained $\rho_b / \rho_r \simeq 8$ (red dot in the figure 9b) and the band gap amplitude $\Delta \simeq 0.5$ (red dot in the figure 9a).

The figures 8 and 9 can be used as a design alternative of tetrachiral materials for given pairs of the band gap amplitude and the center frequency. The table 2 reported two models in which the parameters (32) are obtained respectively from figures 8 and 9, setting $\Delta = 0.45$ and $\gamma = 1.4$.

The effects of an acoustic filter on the propagation of harmonic waves can be observed by modeling a finite dimension system and imposing a harmonic displacement, with frequency $\varpi$, on a boundary [27]. The system is made by two strips of homogeneous material and a core of a tetrachiral material (figure 10). For this example, the parameters (32) assume the values reported for the model 3 in the table 2. By evaluating the dispersion spectra of the homogeneous material and the tetrachiral material, a full band gap, with amplitude $\Delta=0.496$ and center frequency $\gamma=1.675$, is observed in the dispersion spectrum of the tetrachiral material (figure 10).

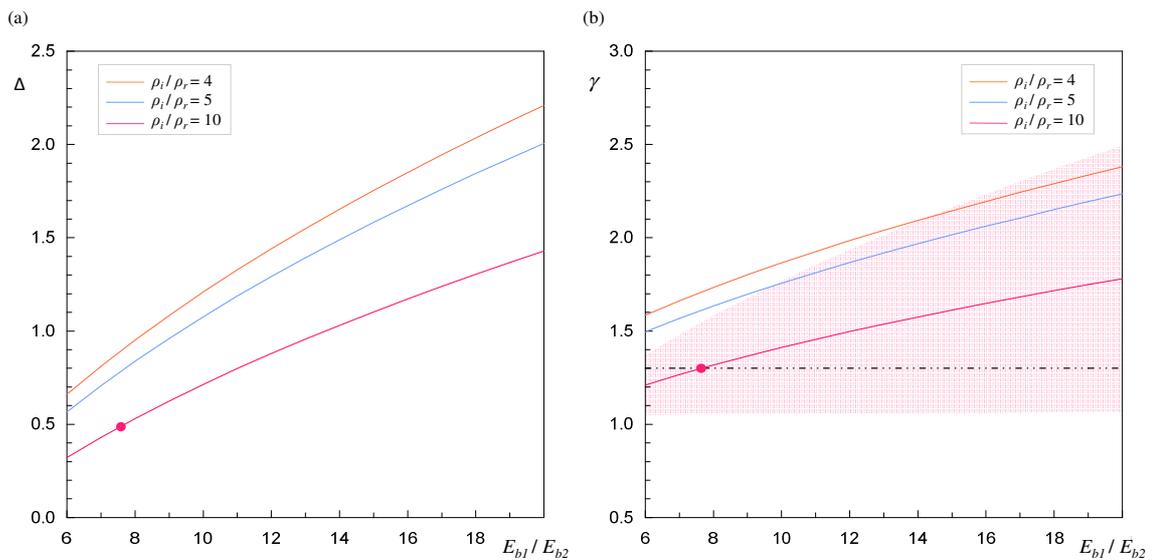

**Figure 9**. (a) Band gap amplitude, (b) Center frequency for a tetrachiral material with inhomogeneous ligaments and intra-ring filler.

Table 2: Parametric values in the models

| | $\rho_i/\rho_r$ | $\rho_b/\rho_r$ | $E_{b1}/E_{b2}$ | $\Delta$ | $\gamma$ |
|---|---|---|---|---|---|
| Model 1 | 10 | 13.5 | 1 | 0.5 | 1.4 |
| Model 2 | 10 | 0 | 9.5 | 0.5 | 1.4 |
| Model 3 | 0 | 10 | 1 | 0.496 | 1.675 |

Imposing to the left boundary a harmonic displacement with frequency $\varpi_1=0.6$ (blue line in the figure 10) a propagation through the tetrachiral core is observed in the figure 11a, where the color map indicates the absolute displacements $D$ at a certain time instant ($t=0.001\ s$). Furthermore, the propagation can also be observed from the time histories of the absolute displacements of the core rings (figure 12a). Indeed, the peak displacement values of the ring 1 (the closest to the excited boundary) are quantitatively comparable with those of the ring 11 (the farthest from the excited boundary). Instead, imposing to the left boundary a harmonic displacement with frequency $\varpi_2=1.6$ (green line in the figure 10), the propagation through the tetrachiral core is stopped (figure 11b). This behavior is due to the value of the excitation frequency, which falls within the band gap of the material spectrum. In particular, the time histories of the material response show that the peaks of the absolute displacements decrease with the distance of the rings from the harmonically excited boundary (figure 12b).

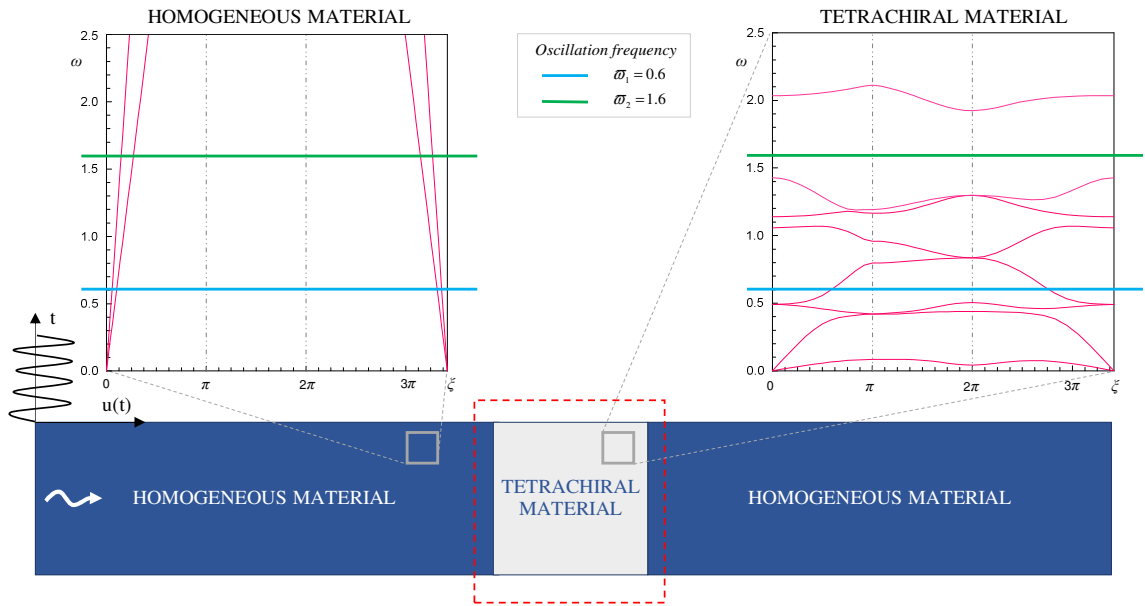

**Figure 10**. Dispersion spectra of the homogeneous material and the tetrachiral material that compose the finite dimension system.

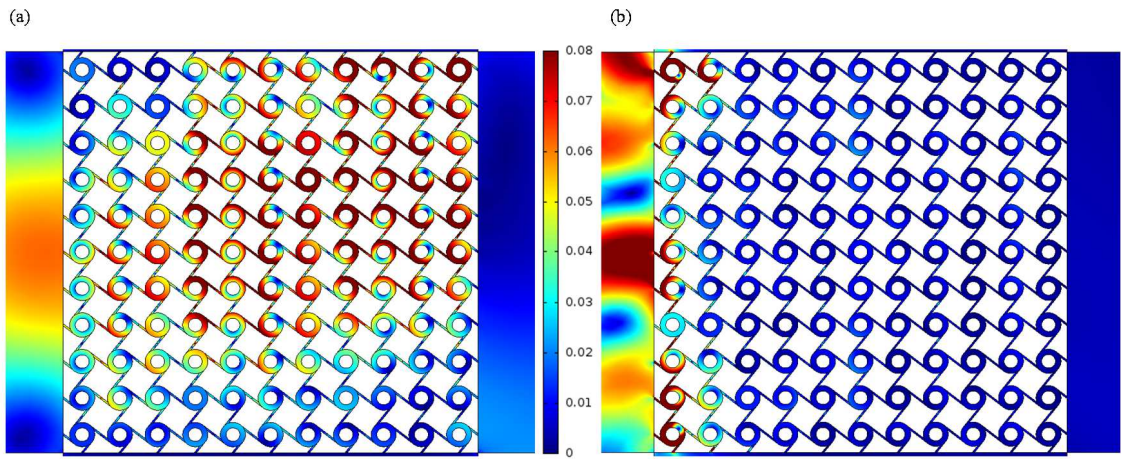

**Figure 11**. Absolute displacements $D$ of the tetrachiral core at the time $t = 0.001\ s$ for an imposed harmonic displacement with frequency (a) $\varpi_1 = 0.6$, (b) $\varpi_2 = 1.6$

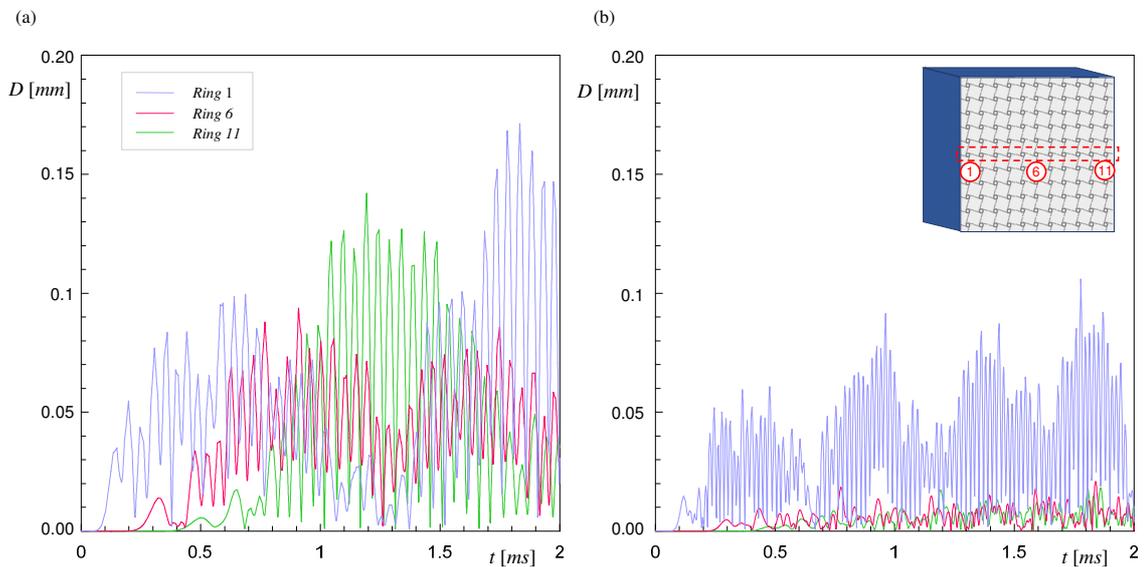

**Figure 12**. Time histories of the absolute displacements $D$ of the ring 1, 6 and 11 of the tetrachiral core for an imposed harmonic displacement with frequency (a) $\varpi_1 = 0.6$, (b) $\varpi_2 = 1.6$.

## 5. Conclusions

The propagation of the elastic waves through a tetrachiral material has been studied using three different models (beam lattice model, first order continuum model and homogenized micropolar continuum model) and a good agreement has been obtained by the comparison of the dispersion spectrum.

Parametric analyses of the dispersion spectrum for a periodic tetrachiral material have been carried out using a solid Cauchy model for different values of geometric and mechanic parameters, with focus on how these differences affect the dispersion surfaces. The analyzes obtained by varying the geometric parameters highlight some issues in satisfying the conditions of the existence of a full band gap in the low-frequency range. These conditions are related to the inertial characteristics of the ring and the slenderness of the ligaments and the physical realization would require a technical arrangement like a material with functionally-graded elastic properties with tapered cross section or other modifications that destroy the invariant properties in the out-of-plane direction. It is possible to obtain a total band gap at a designated frequency using metamaterials with inertial resonators. Alternatively, the possibility to achieve a band gap at a target frequency has been analyzed enlarging the parameter space or removing some hypotheses of the reference model.

In some cases, an enrichment of the low-frequency range with new dispersion surfaces has been observed. For a cell with massive ligaments, these surfaces are related to local waveforms, participating essentially by the ligaments dynamics. Furthermore, increasing the mass density of the ligaments it is possible to obtain, between the optic surfaces, a total band gap in the low-frequency range around the target center frequency.

The opening of a full band gap has been observed also for a tetrachiral material with an intra-ring heavy filler and ligaments made by two material with different Young's Modulus. Increasing the ratio between the Young's Moduli of the ligaments the frequencies of the optic surface grow up and increasing the mass density of the intra-ring filler a total band gap between the second acoustic surface and the first optic surface in the low-frequency range has been observed.

Furthermore, the variation of the mechanic parameters causes changes on the band gap amplitude and on the center frequency and, therefore, the controlled variation of these parameters has been considered a mechanical tuning of the acoustic filter realized by tetrachiral material.

**Acknowledgments**

The authors acknowledge financial support of the (MURST) Italian Department for University and Scientific and Technological Research in the framework of the research MIUR Prin15 project 2015LYYXA8, Multi-scale mechanical models for the design and optimization of micro-structured smart materials and metamaterials, coordinated by prof. A. Corigliano.

**Appendix. Mass and stiffness matrices**

The non-null coefficients $M_{ii}$ $(i=1,2,3)$ of the mass submatrix $\mathbf{M}$ in the equation (1) are

$$M_{11} = M_{22} = M_r \tag{A.1}$$

$$M_{33} = J_r$$

The non-null coefficients $K_{ij}^{aa}$ $(i, j = 1, 2, 3)$ and $K_{hs}^{pp}$ $(h, s = 1,...,12)$ of the symmetric submatrices $\mathbf{K}_{aa}$ and $\mathbf{K}_{pp}$ in the equation (1) are

$$K_{11}^{aa} = K_{22}^{aa} = 4E_b t_b (d^2 H^2 + 4t_b^2) d^{-3} H^{-3} \tag{A.2}$$

$$K_{11}^{aa} = K_{22}^{aa} = 4E_b t_b (d^2 H^2 + 4t_b^2) d^{-3} H^{-3}$$

$$K_{33}^{aa} = \frac{8}{3} E_b t_b (3R^2 + t_b^2) d^{-1} H^{-1}$$

$$K_{11}^{pp} = K_{55}^{pp} = K_{77}^{pp} = K_{1111}^{pp} = 2E_b t_b (d^4 H^4 + 16R^2 t_b^2) d^{-3} H^{-5}$$

$$K_{12}^{pp} = K_{78}^{pp} = -4E_b t_b R \left( H^2 - 4t_b^2 \right) d^{-2} H^{-4}$$

$$K_{45}^{pp} = K_{1011}^{pp} = 4E_b t_b R \left( H^2 - 4t_b^2 \right) d^{-2} H^{-4}$$

$$K_{13}^{pp} = K_{56}^{pp} = 4E_b t_b^3 R d^{-2} H^{-3}$$

$$K_{79}^{pp} = K_{1112}^{pp} = -4E_b t_b^3 R d^{-2} H^{-3}$$

$$K_{22}^{pp} = K_{44}^{pp} = K_{88}^{pp} = K_{1010}^{pp} = 8E_b t_b (R^2 + t_b^2) d^{-1} H^{-3}$$

$$K_{23}^{pp} = K_{1012}^{pp} = 2E_b t_b^3 d^{-1} H^{-2}$$

$$K_{46}^{pp} = K_{89}^{pp} = -2E_b t_b^3 d^{-1} H^{-2}$$

$$K_{33}^{pp} = K_{66}^{pp} = K_{99}^{pp} = K_{1212}^{pp} = \frac{2}{3} E_b t_b^3 d^{-1} H^{-1}$$

where $d = \left(1 - 4R^2 / H^2\right)^{1/2}$.

The coefficients $K_{ij}^{ap}$ $(i=1,2,3; j=1,...,12)$ of the rectangular submatrix $\mathbf{K}_{ap} = \mathbf{K}_{pa}^T$ in the equation (1) are

$$K_{11}^{ap} = K_{17}^{ap} = K_{25}^{ap} = K_{211}^{ap} = -2E_b t_b (d^4 H^4 + 16R^2 t_b^2) d^{-3} H^{-5} \quad (A.3)$$

$$K_{12}^{ap} = K_{21}^{ap} = K_{18}^{ap} = K_{27}^{ap} = 4E_b t_b R\left(H^2 - 4\left(R^2 + t_b^2\right)\right) d^{-2} H^{-4}$$

$$K_{15}^{ap} = K_{111}^{ap} = K_{24}^{ap} = K_{210}^{ap} = -4E_b t_b R\left(H^2 - 4\left(R^2 + t_b^2\right)\right) d^{-2} H^{-4}$$

$$K_{13}^{ap} = K_{26}^{ap} = -4E_b t_b^3 R d^{-2} H^{-3}$$

$$K_{19}^{ap} = K_{212}^{ap} = 4E_b t_b^3 R d^{-2} H^{-3}$$

$$K_{14}^{ap} = K_{110}^{ap} = K_{22}^{ap} = K_{28}^{ap} = -8E_b t_b (R^2 + t_b^2) d^{-1} H^{-3}$$

$$K_{16}^{ap} = K_{29}^{ap} = 2E_b t_b^3 d^{-1} H^{-2}$$

$$K_{112}^{ap} = K_{23}^{ap} = -2E_b t_b^3 d^{-1} H^{-2}$$

$$K_{31}^{ap} = K_{35}^{ap} = -2E_b t_b R\left(H^2 - 2\left(2R^2 + t_b^2\right)\right) d^{-2} H^{-3}$$

$$K_{37}^{ap} = K_{311}^{ap} = 2E_b t_b R\left(H^2 - 2\left(2R^2 + t_b^2\right)\right) d^{-2} H^{-3}$$

$$K_{32}^{ap} = K_{310}^{ap} = 2E_b t_b (2R^2 + t_b^2) d^{-1} H^{-2}$$

$$K_{34}^{ap} = K_{38}^{ap} = -2E_b t_b (2R^2 + t_b^2) d^{-1} H^{-2}$$

$$K_{33}^{ap} = K_{36}^{ap} = K_{39}^{ap} = K_{312}^{ap} = \frac{1}{3} E_b t_b^3 d^{-1} H^{-1}$$